\documentclass[english,aps,prc,twocolumn,showpacs,floatfix]{revtex4}
\usepackage[T1]{fontenc}
\usepackage[latin1]{inputenc}
\usepackage{graphicx}
\usepackage{babel}
\newcommand{\ba}{\begin{eqnarray}}
\newcommand{\ea}{\end{eqnarray}}

\usepackage{epsfig}
\usepackage{graphics}

\begin{document}

\title{Unquenched quark model for baryons: magnetic moments, spins and orbital angular momenta}

\author{ R. Bijker}
\affiliation{Instituto de Ciencias Nucleares, 
Universidad Nacional Aut\'onoma de M\'exico, 
A.P. 70-543, 04510 M\'exico, D.F., M\'exico}
\email{bijker@nucleares.unam.mx}

\author{E. Santopinto}
\affiliation{I.N.F.N., Sezione di Genova, via Dodecaneso 33, I-16146 Italy}
\email{santopinto@ge.infn.it}

\date{\today}

\begin{abstract} 
We present an unquenched quark model for baryons in which the effects of the 
quark-antiquark pairs ($u \bar{u}$, $d \bar{d}$ and $s \bar{s}$) are taken 
into account in an explicit form via a microscopic, QCD-inspired, quark-antiquark 
creation mechanism. In the present approach, the contribution of the quark-antiquark 
pairs can be studied for any inital baryon and for any flavor of the $q \bar{q}$ 
pairs. It is shown that, while the inclusion of the $q \bar{q}$ pairs does not affect 
the baryon magnetic moments, it leads to a sizeable contribution of the orbital 
angular momentum to the spin of the proton and the $\Lambda$ hyperon. 
\end{abstract}

\pacs{12.39.-x, 14.65.Bt, 14.20.Dh, 14.20.Jn}

\maketitle

\section{Introduction}

One of the main goals of hadronic physics is to understand the structure of the 
nucleon and its excited states in terms of effective degrees of freedom and, 
at a more fundamental level, the emergence of these effective degrees of freedom 
from QCD, the underlying theory of quarks and gluons \cite{Isgur}. 
Despite the progress made in lattice calculations, it remains a daunting problem 
to solve the QCD equations in the non-perturbative region. Therefore, one has 
developed effective models of hadrons, such as bag models, chiral quark models, 
soliton models \cite{cbm}, instanton liquid model \cite{shuryak} and the constituent 
quark model. Each of these approaches is constructed in order to mimic some 
selected properties of the strong interaction, but obviously none of them is QCD. 

\begin{figure}[tb]
\centering
\resizebox{0.45\textwidth}{!}{\includegraphics{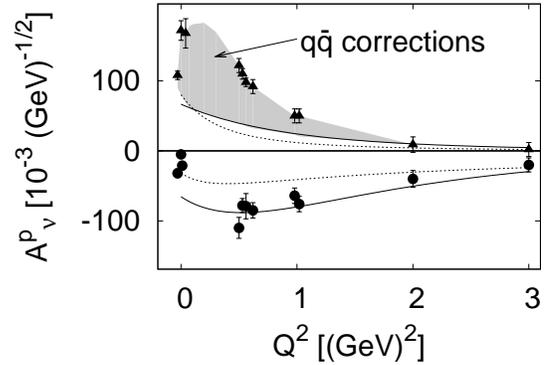}}
\caption[]{The helicity amplitudes as a function of $Q^2$ for the $D_{13}(1520)$ resonance. 
Experimental data \cite{d13_exp} are compared with theoretical predictions from the collective 
$U(7)$ model \cite{bil} (dotted line) and the hypercentral model \cite{hypercentral} 
(solid line). The dashed line corresponds to a fit to the experimental data. }
\label{d13}
\end{figure}

An important class is provided by constituent quark models (CQM) which are  
based on constituent (effective) quark degrees of freedoms. There exists a large 
variety of CQMs, among others the Isgur-Karl model \cite{ik}, the Capstick-Isgur 
model \cite{ci}, the collective model \cite{bil}, the hypercentral model 
\cite{hypercentral}, the chiral boson-exchange model \cite{olof} and the Bonn 
instanton model \cite{bn}. While these models display important and peculiar 
differences, they share the main features: the effective degrees of freedom of three 
constituent quarks ($qqq$ configurations), the $SU(6)$ spin-flavor symmetry 
and a long-range confining potential. Each of these models reproduce the mass 
spectrum of baryon resonances reasonably well, but at the same time, they show 
very similar deviations for other observables, such as photocouplings, helicity 
amplitudes and strong decays. As an example, we mention helicity amplitudes 
(or transition form factors) which typically show deviations from CQM calculations 
at low values of $Q^2$ (see Fig.~\ref{d13} for the $D_{13}(1520)$ resonance). 
The problem of missing strength at low $Q^2$ in constituent quark model calculations 
indicates that some fundamental mechanism is lacking in the dynamical description 
of hadronic structure. This mechanism can be identified with the production of 
quark-antiquark pairs \cite{aie,lothar}. Low values of $Q^2$ correspond to a 
distance scale at which there is a higher probability of string breaking and thus 
of quark-antiquark pair production. 

Additional evidence for higher Fock components in the baryon wave function 
($qqq-q\bar{q}$ configurations) comes from CQM studies of the strong decays of baryon 
resonances, that are on average underpredicted by CQMs \cite{bil,CR}.
More direct indications for the importance of quark-antiquark components in the proton 
come from measurements of the $\bar{d}/\bar{u}$ asymmetry in the nucleon sea 
\cite{Kumano,Garvey} and parity-violating electron scattering experiments  
which report a nonvanishing strange quark contribution, albeit (very) small, to 
the charge and magnetization distributions \cite{Acha,Baunack}. 

The role of higher Fock components in baryon wave functions has been studied by many authors  
in the context of meson cloud models, such as the cloudy bag model, meson convolution models 
and chiral models \cite{Kumano,Speth}. In these models, the flavor asymmetry of the proton 
can be understood in terms of couplings to the pion cloud. 
There have also been several attempts to study the importance of higher Fock components 
in the context of the constituent quark model. In this respect we mention the work by Riska 
and coworkers who introduce a small number of selected higher Fock components which are 
then fitted to reproduce the experimental data \cite{riska}. However, these studies 
lack an explicit model or mechanism for the mixing between the valence and sea quarks. 
The Rome group studied the pion and nucleon electromagnetic form factors in a Bethe-Salpeter 
approach, mainly thanks to the dressing of photon vertex by means of a vector-meson dominance 
parametrization \cite{rm_pi}. 
Koniuk and Guiasu used a convolution model with CQM wave functions and an elementary emission 
model for the coupling to the pion cloud to calculate the magnetic moments and the helicity 
amplitudes from the nucleon to the $\Delta$ resonance \cite{convolution}. It was found that 
the nucleon magnetic moments were unchanged after renomalization of the parameters, but that 
the missing strength in the helicity amplitudes of the $\Delta$ could not be explained with 
pions only.

\begin{figure}[tb]
\centering
\resizebox{0.45\textwidth}{!}{\includegraphics{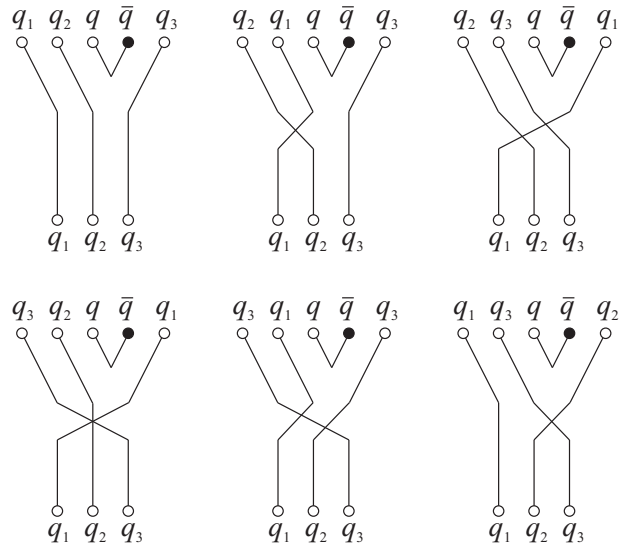}}
\vspace{15pt}
\caption[]{\small Quark line diagrams for $A \rightarrow B C$ 
with $q\bar{q}=s \bar{s}$ and $q_1q_2q_3=uud$.}
\label{diagrams}
\end{figure}

The impact of $q\bar{q}$ pairs in hadron spectroscopy was originally studied by T\"ornqvist 
and Zenczykowski in a quark model extended by the $^{3}P_0$ model \cite{tornqvist}. Even though 
their model only includes a sum over ground state baryons and ground state mesons, the basic 
idea of the importance to carry out a sum over a complete set of intermediate states was 
proposed in there.  
Subsequently, the effects of hadron loops in mesons was studied by Geiger and Isgur in 
a flux-tube breaking model in which the $q\bar{q}$ pairs are created in the $^{3}P_0$ state 
with the quantum numbers of the vacuum \cite{kokoski,geiger,OZI}. In this approach, the quark 
potential model arises from an adiabatic approximation to the gluonic degrees of freedom 
embodied in the flux-tube \cite{paton}. It was shown that cancellations between apparently 
uncorrelated sets of intermediate states occur in such a way that the modification in the 
linear potential can be reabsorbed, after renormalization, in the new strength of the linear 
potential \cite{geiger}. In addition, the quark-antiquark pairs do not destroy the good 
CQM results for the mesons \cite{geiger} and preserve the OZI hierarchy \cite{OZI} 
provided that the sum be carried out over a large tower of intermediate states. 
A first application of this procedure to baryons was presented in \cite{baryons} in which 
the importance of $s \bar{s}$ loops in the proton were studied by taking into account the 
contribution of the six diagrams of Fig.~\ref{diagrams} in combination with harmonic 
oscillator wave functions for the baryons and mesons and a $^3 P_0$ pair creation mechanism. 
This approach has the advantage that the effects of quark-antiquark pairs are introduced 
explicitly via a QCD-inspired pair-creation mechanism, which opens the possibility to study 
the importance of $q \bar{q}$ pairs in baryons and mesons in a systematic and unified way. 

The aim of this article is to present an unquenched quark model, valid for any initial baryon 
(or baryon resonance), any flavor of the quark-antiquark pair (not only $s \bar{s}$, but also 
$u \bar{u}$ and $d \bar{d}$ loops) and any CQM. 
In order to test the consistency of the formalism we first calculate the baryon magnetic moments 
which constitute one of the early successes of the CQM. Finally, we study an application of the 
unquenched quark model to the spin of the proton and the $\Lambda$ hyperon, and calculate in 
explicit form the contributions of the valence and sea quark spins and the orbital angular 
momentum. Preliminary results of this work were presented in various conference proceedings 
\cite{ucqm1,ucqm2,ucqm3}.

\section{Unquenched quark model \label{ucqm}}

In this section, we present a procedure for unquenching the quark model in which the effects of 
quark-antiquark pairs are introduced explicitly into the CQM via a QCD-inspired $^{3}P_0$ 
pair-creation mechanism. The present approach is motivated by the work of Isgur and coworkers 
on the flux-tube breaking model in which they showed that the CQM emerges as the adiabatic 
limit of the flux-tube model to which the effects of $q \bar{q}$ pair creation are added 
as a perturbation \cite{baryons}. Our approach is based on a CQM to which the quark-antiquark 
pairs with vacuum quantum numbers are added as a perturbation. The pair-creation mechanism is 
inserted at the quark level and the one-loop diagrams are calculated by summing over all  
possible intermediate states. 

Under these assumptions, the baryon wave function consists of a zeroth order three-quark 
configuration plus a sum over all possible higher Fock components due to the creation of 
$^{3}P_0$ quark-antiquark pairs. To leading order in pair creation, the baryon wave function 
can be written as  
\begin{eqnarray} 
\mid \psi_A \rangle &=& {\cal N} \left[ \mid A \rangle 
+ \sum_{BC l J} \int d \vec{k} \, \mid BC \vec{k} \, l J \rangle \right.
\nonumber\\
&& \hspace{2cm} \left. \times 
\frac{ \langle BC \vec{k} \, l J \mid T^{\dagger} \mid A \rangle } 
{M_A - E_B - E_C} \right] ~,
\label{baryonwf} 
\end{eqnarray}
where $T^{\dagger}$ is the $^{3}P_0$ quark-antiquark pair creation operator 
\cite{roberts}, $A$ is the baryon, $B$ and $C$ represent the intermediate baryon and meson, 
and $M_A$, $E_B$ and  $E_C$ are  their respective energies, 
$\vec{k}$ and $l$ the relative radial momentum and orbital angular 
momentum of $B$ and $C$, and $J$ is the total angular momentum 
$\vec{J} = \vec{J}_B + \vec{J}_C + \vec{l}$. 

The $^{3}P_0$ quark-antiquark pair-creation operator, $T^{\dagger}$, can be written as  \cite{roberts}
\begin{eqnarray}
T^{\dagger} &=& -3 \, \gamma_0 \,\int d \vec{p}_4 \, d \vec{p}_5 \, 
\delta(\vec{p}_4 + \vec{p}_5) \, C_{45} \, F_{45} \,  
{e}^{-r_q^2 (\vec{p}_4 - \vec{p}_5)^2/6 }\, 
\nonumber\\
&& \hspace{1cm} \left[ \chi_{45} \, \times \, {\cal Y}_{1}(\vec{p}_4 - \vec{p}_5) \right]^{(0)}_0 \, 
b_4^{\dagger}(\vec{p}_4) \, d_5^{\dagger}(\vec{p}_5) ~.   
\label{3p0}
\end{eqnarray}
Here, $b_4^{\dagger}(\vec{p}_4)$ and $d_5^{\dagger}(\vec{p}_5)$ are the creation operators 
for a quark and an antiquark with momenta $\vec{p}_4$ and $\vec{p}_5$, respectively. The 
quark and antiquark pair is characterized by a color singlet wave function $C_{45}$, a 
flavor singlet wave function $F_{45}$, a spin triplet wave function $\chi_{45}$ with spin 
$S=1$ and a solid spherical harmonic ${\cal Y}_{1}(\vec{p}_4 - \vec{p}_5)$ that indicates that 
the  quark and antiquark are in a relative $P$ wave. The operator $T^{\dagger}$ creates a pair 
of constituent quarks with an effective size, thus the pair creation point is smeared
out by a gaussian factor whose width $r_q$ was determined from meson decays to be 
approximately $0.25-0.35$ fm \cite{baryons,OZI,SBG}. In our calculations, we take an average 
value, $r_q=0.30$ fm. The dimensionless constant $\gamma_0$ is the intrinsic pair-creation 
strength which was determined from strong decays of baryons as $\gamma_0=2.60$ \cite{CR}. 
The matrix elements of the pair-creation operator $T^{\dagger}$ were derived in explicit form 
in the harmonic oscillator basis \cite{roberts}. 

In this paper, we use the harmonic oscillator limit of algebraic models of hadron structure 
\cite{bil,algebraic} to calculate the baryon and meson energies appearing in the denominator 
of Eq.~(\ref{baryonwf}). In these algebraic models, the mass operators for baryons and mesons 
consist of a harmonic oscillator term and a G\"ursey-Radicati term which reproduces the splitting 
of the $SU(6)$ multiplets without mixing the harmonic oscillator wave functions. 
As a consequence, the baryon and meson wave functions have good flavor symmetry and depend on a single 
oscillator parameter which, following \cite{baryons}, is taken to be $\hbar \omega_{\rm baryon}=0.32$ 
GeV for the baryons and $\hbar \omega_{\rm meson}=0.40$ GeV for the mesons. 

The matrix elements of an observable $\hat{\cal O}$ can be calculated as 
\begin{eqnarray} 
{\cal O} = \langle \psi_A \mid \hat{\cal O} \mid \psi_A \rangle 
= {\cal O}_{\rm valence} + {\cal O}_{\rm sea} ~,
\end{eqnarray}  
where the first term corresponds to the contribution of the three valence quarks 
and the second to the higher Fock components, {\it i.e.} the presence of the quark-antiquark pairs. 

In order to calculate the effects of quark-antiquark pairs on an observable, one has to 
evaluate the sum over all possible intermediate states in Eq.~(\ref{baryonwf}). 
The sum over intermediate meson-baryon states includes for baryons all radial and orbital 
exications up to a given oscillator shell combined with all possible $SU(6)$ spin-flavor 
multiplets, and for mesons all radial and orbital excitations up to given oscillator shell and all 
possible nonets. 
This problem was solved by means of group theoretical techniques to construct an algorithm to 
generate a complete set of intermediate meson-baryon states in spin-flavor space for an arbitrary 
oscillator shell. This property makes it possible to perform the sum over intermediate states up 
to saturation and not only for the first few shells as in \cite{baryons}. In addition, it allows 
the evaluation of the contribution of quark-antiquark pairs for any initial baryon $q_1 q_2 q_3$ 
(ground state or resonance) and for any flavor of the $q \bar{q}$ pair (not only $s\bar{s}$, 
but also $u\bar{u}$ and $d\bar{d}$), and for any model of baryons and mesons, as long as their 
wave functions are expressed in the basis of harmonic oscillator wave functions. 

\section{Closure limit}

Before discussing an application of the unquenched model to baryon magnetic moments and spins, 
we study the so-called closure limit in which the intermediate states appearing 
in Eq.~(\ref{baryonwf}) are degenerate in energy and hence the energy denominator becomes a constant 
independent of the quantum numbers of the intermediate states. In the closure limit, the evaluation 
of the contribution of the quark-antiquark pairs (or the higher Fock components) simplifies considerably, 
since the sum over intermediate states can be solved by closure and the contribution 
of the quark-antiquark pairs to the matrix element reduces to 
\begin{eqnarray}
{\cal O}_{\rm sea} &\propto& \langle A \mid T \, \hat{\cal O} \,   
T^{\dagger} \mid A \rangle ~.
\end{eqnarray}
Since the $^{3}P_0$ pair-creation operator of Eq.~(\ref{3p0}) is a flavor singlet and the 
energy denominator in Eq.~(\ref{baryonwf}) is reduced to a constant in the closure limit, 
the higher Fock component of the baryon wave function has the same flavor symmetry as the 
valence quark configuration $\mid A \rangle$. 
Moreover, if the pair-creation operator does not couple to the motion of the valence quarks, 
the valence quarks act as spectators. In this case, the contribution of the $q \bar{q}$ pairs 
simplifies  further to the expectation value of $\hat{\cal O}$ between the $^{3}P_0$ pair 
states created by $T^{\dagger}$ 
\begin{eqnarray}
{\cal O}_{\rm sea} &\propto& \langle 0 \mid T \, \hat{\cal O} \,   
T^{\dagger} \mid 0 \rangle ~, 
\end{eqnarray}
the so-called closure-spectator limit \cite{baryons} which is a special case of the closure limit.  

\begin{table}[tb]
\centering
\caption[]{$\Delta u$, $\Delta d$ and $\Delta s$ for ground state octet baryons in the closure limit 
in units of $(\Delta u)_p/4$.}
\label{spinoct}
\begin{tabular}{ccrcrcr}
\noalign{\smallskip}
\hline
\noalign{\smallskip}
$qqq$ & $^{2}8[56,0^+]$  & $\Delta u$ && $\Delta d$ && $\Delta s$ \\
\noalign{\smallskip}
\hline
\noalign{\smallskip}
$uud$ & $p$ & 4 && $-1$ && 0 \\
$udd$ & $n$ & $-1$ && 4 && 0 \\
$uus$ & $\Sigma^+$ & 4 && 0 && $-1$ \\
$uds$ & $\Sigma^0$ & 2 && 2 && $-1$ \\
      & $\Lambda$  & 0 && 0 && 3 \\
$dds$ & $\Sigma^-$ & 0 && 4 && $-1$ \\
$uss$ & $\Xi^0$ & $-1$ && 0 && 4 \\
$dss$ & $\Xi^-$ & 0 && $-1$ && 4 \\
\noalign{\smallskip}
\hline
\end{tabular}
\end{table}

As an example, we discuss the contribution of the quark-antiquark pairs for the operator 
$2[s_z(q) + s_z(\bar{q})]$ in the closure limit 
\begin{eqnarray}
\Delta q = 2 \left< s_z(q) + s_z(\bar{q}) \right> ~.
\end{eqnarray}
$\Delta q$ is the nonrelativistic limit of the axial charges and denotes the fraction of 
the baryon's spin carried by quarks and antiquarks with flavor $q=u$, $d$, $s$. In 
Table~\ref{spinoct} we present the results for the ground state octet baryons with 
$^{2}8[56,0^+]_{1/2}$. Since the valence-quark configuration of the proton and the neutron 
does not contain strange quarks, the valence quarks act as spectators in the calculation of 
$\Delta s$. Therefore, the contribution of $\Delta s$ to the spin of the nucleon is given by 
the closure-spectator limit which vanishes due to the symmetry properties of the operator 
$\Delta s$ and the $^{3}P_0$ wave function. The same holds for the contribution 
of $d \bar{d}$ pairs to the $\Sigma^{+}$ and $\Xi^{0}$ hyperons, and that of $u \bar{u}$ 
pairs to the $\Sigma^{-}$ and $\Xi^{-}$ hyperons. The vanishing contributions of $\Delta u$ 
and $\Delta d$ to the spin of the $\Lambda$ hyperon are a consequence of the $\Lambda$ wave 
function in which the up and down quarks are coupled to isospin and spin zero. Similarly, 
the vanishing contributions of $\Delta q$ to the spin of the ground state decuplet baryons with 
$^{4}10 [56,0^+]_{3/2}$ in Table~\ref{spindec} can be understood in the closure-spectator limit. 

In addition, since in the closure limit the baryon wave function has the same flavor symmetry 
as the valence quark configuration, it can be shown that the flavor dependence of the 
contribution of the quark-antiquark pairs to the spin of the ground state baryons in 
Tables~\ref{spinoct} and \ref{spindec} is the same as that of the valence quarks 
\begin{eqnarray}
\Delta u_{\rm sea} : \Delta d_{\rm sea} : \Delta s_{\rm sea} = 
\Delta u_{\rm val} : \Delta d_{\rm val} : \Delta s_{\rm val} ~.
\label{seaval}
\end{eqnarray}
The results for octet and decuplet ground state baryons are related by
\begin{eqnarray}
\left( \Delta u + \Delta d + \Delta s \right)_{\rm dec} = 
3\left( \Delta u + \Delta d + \Delta s \right)_{\rm oct} ~.
\end{eqnarray}
The same relation holds for the orbital angular momentum 
\begin{eqnarray}
\left( \Delta L \right)_{\rm dec} = 3 \left( \Delta L \right)_{\rm oct} ~, 
\end{eqnarray}
with
\begin{eqnarray}
\Delta L = \sum_q \Delta L(q) = \sum_q \left< l_z(q) + l_z(\bar{q}) \right> ~.
\end{eqnarray}
Note that, even if the valence quark configuration $[56,0^+]$ does not carry orbital angular 
momentum, there is a nonzero contribution of the quark-antiquark pairs in the closure limit, 
albeit small (less than 1 \%) in comparison with that of the quark spins. Obviously, the sum 
of the spin and orbital parts gives the total angular momentum of the baryon 
\begin{eqnarray}
J = \frac{1}{2} \Delta \Sigma + \Delta L ~,
\end{eqnarray}
with
\begin{eqnarray}
\Delta \Sigma = \Delta u + \Delta d + \Delta s ~.
\end{eqnarray}

\begin{table}[tb]
\centering
\caption[]{As Table~\ref{spinoct}, but for ground state decuplet baryons.}
\label{spindec}
\begin{tabular}{ccccccc}
\noalign{\smallskip}
\hline
\noalign{\smallskip}
$qqq$ & $^{4}10[56,0^+]$ & $\Delta u$ & & $\Delta d$ & & $\Delta s$ \\
\noalign{\smallskip}
\hline
\noalign{\smallskip}
$uuu$ & $\Delta^{++}$ & 9 & & 0 & & 0 \\
$uud$ & $\Delta^{+}$ & 6 & & 3 & & 0 \\
$udd$ & $\Delta^{0}$ & 3 & & 6 & & 0 \\
$ddd$ & $\Delta^{-}$ & 0 & & 9 & & 0 \\
$uus$ & $\Sigma^{\ast \, +}$ & 6 & & 0 & & 3 \\
$uds$ & $\Sigma^{\ast \, 0}$ & 3 & & 3 & & 3 \\
$dds$ & $\Sigma^{\ast \, -}$ & 0 & & 6 & & 3 \\
$uss$ & $\Xi^{\ast \, 0}$ & 3 & & 0 & & 6 \\
$dss$ & $\Xi^{\ast \, -}$ & 0 & & 3 & & 6 \\
$sss$ & $\Omega^{-}$ & 0 & & 0 & & 9 \\
\noalign{\smallskip}
\hline
\end{tabular}
\end{table}

At a qualitative level, the closure limit helps to explain the phenomenological success 
of the CQM because the $SU(3)$ flavor symmetry of the baryon wave function is preserved. 
As an example, the strange content of the proton vanishes in the closure-spectator limit 
due to many cancelling contributions in the sum over intermediate states in Eq.~(\ref{baryonwf}). 
Away from the closure limit, the strangeness content of the proton is expected to be 
small, in agreement with the experimental data from parity-violating electron scattering 
(for some recent data see \cite{Acha,Baunack}). Even though in this case the cancellations 
are no longer exact, many intermediate states contribute with opposite signs, and the net 
result is nonzero, but small. This means that even if the flavor symmetry of the CQM is 
broken by the higher Fock components, the net results are still to a large extent determined 
by the flavor symmetry of the valence quark configuration. Similar arguments were applied to 
the preservation of the OZI hierarchy in the context of the flux-tube breaking model \cite{OZI}. 
Therefore, the closure limit not only provides simple expressions for the relative flavor 
content of physical observables, but also gives further insight into the origin of 
cancellations between the contributions from different intermediate states. 

In addition, the results in closure limit in Tables~\ref{spinoct} and \ref{spindec} impose 
very stringent conditions on the numerical calculations, since each entry involves the sum 
over all possible intermediate states. Therefore, the closure limit provides a highly 
nontrivial test of the computer codes which involves both the spin-flavor sector, the 
permutation symmetry, the construction of a complete set of intermediate states in spin-flavor 
space for each radial excitation and the implementation of the sum over all of these states. 

In this section, we discussed some qualitative properties of the unquenched quark model 
in the closure limit. In the following sections, we study the effects of quark-antiquark 
pairs on the magnetic moments and the spin of octet baryons in the general case, 
{\it i.e.} beyond the closure limit. 

\section{Magnetic moments}

The unquenching of the quark model has to be carried out in such a way as to preserve the 
phenomenological successes of the constituent quark model. It is well known that the CQM 
gives a good description of the baryon magnetic moments, even in its simplest form in which 
the baryons are treated in terms of three constituent quarks in a relative $S$-wave. 
The quark magnetic moments are determined by fitting the magnetic moments of the proton, 
neutron and $\Lambda$ hyperon to give $\mu_u=1.852$, $\mu_d=-0.972$ and $\mu_s=-0.613$ 
$\mu_N$ \cite{pdg}. 

\begin{figure}[tb]
\centering
\resizebox{0.45\textwidth}{!}{\includegraphics{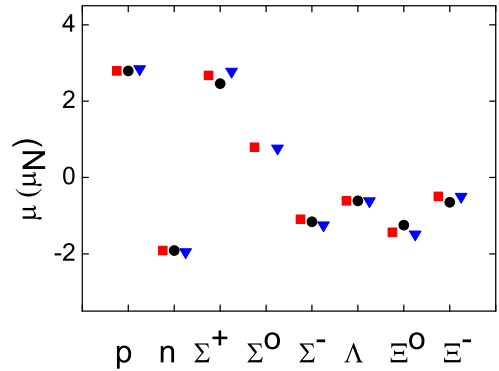}}
\caption[]{(color online) Magnetic moments of octet baryons: experimental values from PDG 
\cite{pdg} (circles), CQM (squares) and unquenched quark model (triangles).}
\label{magmom}
\end{figure}

In the unquenched CQM the baryon magnetic moments also receive contributions from the quark 
spins of the pairs and the orbital motion of the quarks 
\ba
\vec{\mu} = \sum_{q} \mu_q \left[ 2 \vec{s}(q) + \vec{l}(q) 
- 2 \vec{s}(\bar{q}) - \vec{l}(\bar{q}) \right] ~,
\label{magnetic}
\ea
where $\mu_q=e_q \hbar/2m_q c$ is the quark magnetic moment. In Fig.~\ref{magmom} we show a  
comparison between the experimental values of the magnetic moments of the octet baryons (circles) 
and the theoretical values obtained in the CQM (squares) and in the unquenched quark model 
(triangles). The results for the unquenched quark model were obtained in a calculation  
involving a sum over intermediate states up to five oscillator shells for both baryons and mesons. 
We note, that the results for the magnetic moments, after renormalization, are almost 
independent on the number of shells included in the sum over intermediate states. 
The values of the magnetic moments in the unquenched 
quark model are very similar to those in the CQM. The largest difference is observed for the 
charged $\Sigma$ hyperons, but the relation between the magnetic moments of $\Sigma$ hyperons 
\cite{CG}, $\mu(\Sigma^0)=[\mu(\Sigma^+)+ \mu(\Sigma^-)]/2$, is preserved in the unquenched 
calculation due to isospin symmetry. 

The inclusion of the $q\bar{q}$ pairs leads to slightly different values of the quark magnetic 
moments, $\mu_u=2.066$, $\mu_d=-1.110$ and $\mu_s=-0.633$ $\mu_N$ as for the CQM. This is 
related to the well-known phenomenon, that a calculation carried out in a truncated basis leads 
to effective parameters in order to reproduce the results obtained in a more extended basis. 
The results in the unquenched quark model are practically identical, after renormalization, to 
the ones in the CQM, which shows that the addition of the quark-antiquark pairs preserves the 
good CQM results for the baryon magnetic moments. A similar feature was found in the context of 
the flux-tube breaking model for mesons in which it was shown that the inclusion of quark-antiquark 
pairs preserved the linear behavior of the confining potential as well as the OZI hierarchy 
\cite{OZI}. The change in the linear potential caused by the bubbling of the pairs in the string 
could be absorbed into a renormalized strength of the linear potential. 

The results for the magnetic moments can be understood qualitatively in the closure limit in 
which the relative contribution of the quark spins from the quark-antiquark pairs is the same 
as that from the valence quarks. Moreover, since in the closure limit the contribution of the 
orbital angular momentum is small in comparison to that of the quark spins, the results for the 
baryon magnetic moments are almost indistinguishable from those of the CQM. 
Away from the closure limit, even though the relations between the different contributions 
no longer hold exactly, they are still valid approximately. In addition, there is now a contribution 
from the orbital part (at the level of $\sim 5$ \%) which is mainly due to the baryon-pion channel. 

In summary, the inclusion of the effects of quark-antiquark pairs preserves, after renormalization, 
the good results of the CQM for the magnetic moments of the octet baryons. 
 
\section{Spins and orbital angular momenta}

In this section, we discuss an application of the unquenched quark model to the spin content of the 
proton and the $\Lambda$ hyperon. 
Ever since the European Muon Collaboration at CERN showed that {\it the total quark spin 
constitutes a rather small fraction of the spin of the nucleon} \cite{emc}, there has been an 
enormous interest in the spin structure of the proton \cite{protonspin,bassreview,leader}. 
The original EMC result suggested that the contribution of the quark spins was close to zero, 
$\Delta \Sigma=0.120 \pm 0.094 \pm 0.138$ \cite{emc}. Thanks to a new generation of experiments 
and an increase in experimental accuracy, the fraction of the proton spin carried by the quarks 
and antiquarks is now known to be about one third. The most recent values were obtained by the 
HERMES and COMPASS collaborations, $\Delta \Sigma = 0.330 \pm 0.011 \pm 0.025 \pm 0.028$ at 
$Q^2=5$ GeV$^2$ \cite{hermes} and $0.33 \pm 0.03 \pm 0.05$ at $Q^2=3$ GeV$^2$ \cite{compass}, 
respectively. The EMC results led to the idea that the proton might contain a substantial amount 
of polarized glue which could contribute to reducing the contribution of the quark spins through 
the $U(1)$ axial anomaly \cite{ef}. Therefore, much of the early theoretical work was in the 
direction of understanding the role of polarized gluons and the axial anomaly to resolve the puzzle 
of the proton spin \cite{ef,anselmino,protonspin}. However, there is increasing evidence from 
recent experiments, that at low values of $Q^2$ the gluon contribution is rather small (either 
positive or negative) and compatible with zero \cite{gluonexp,gluonth}, which rules out the 
possibility that most of the missing spin be carried by the gluon. At the same time, this indicates 
that the missing spin of the proton has to be attributed to others mechanisms 
\cite{bassreview,leader}, in particular to the orbital angular momentum of the quarks and 
antiquarks \cite{Sehgal,orbital,ucqm2,thomas_spin}. 

\subsection{Proton spin}

\begin{table}[tb]
\centering
\caption[]{\small Contribution of $\Delta u$, $\Delta d$ , $\Delta s$, 
$\Delta \Sigma = \Delta u + \Delta d + \Delta s$ and $\Delta L$ to the proton spin 
in the unquenched quark model (UCQM).}
\label{spinproton}
\begin{tabular}{ccrrrrr}
\noalign{\smallskip}
\hline
\noalign{\smallskip}
& & & & \multicolumn{3}{c}{UCQM} \\
$p$ & CQM & EJS & DIS & val & sea & total \\
\noalign{\smallskip}
\hline
\noalign{\smallskip}
$\Delta u$      &   4/3 &   0.928 &   0.842 &   0.504 &   0.594 &   1.098 \\
$\Delta d$      & --1/3 & --0.342 & --0.427 & --0.126 & --0.291 & --0.417 \\
$\Delta s$      &   0   &   0.000 & --0.085 &   0.000 & --0.005 & --0.005 \\
$\Delta \Sigma$ &   1   &   0.586 &   0.330 &   0.378 &   0.298 &   0.676 \\
$2 \Delta L$    &   0   &   0.414 &         &   0.000 &   0.324 &   0.324 \\
$2J$            &   1   &   1.000 &         &   0.378 &   0.622 &   1.000 \\
\noalign{\smallskip}
\hline
\end{tabular}
\end{table}

The formalism developed in Section~\ref{ucqm} makes it possible to study the effect of 
quark-antiquark pairs on the fraction of the proton spin carried by the quark (antiquark) 
spins and orbital angular momentum by means of an explicit calculation in an unquenched quark model.  
Just as in other effective models \cite{bassreview,CBM,NJL} the unquenched quark model 
does not include gluonic effects associated with the axial anomaly, and therefore the contribution 
from the gluons is missing from the outset. The total spin of the proton can be written as
\begin{eqnarray}
\frac{1}{2} = \frac{1}{2} \Delta \Sigma + \Delta L 
= \frac{1}{2} \left( \Delta u + \Delta d + \Delta s \right) + \Delta L ~.
\label{jproton}
\end{eqnarray}
The axial charges,  
\begin{eqnarray}
\Delta q = \left< p \uparrow | \bar{q} \gamma_z \gamma_5 q | p \uparrow \right> ~,
\end{eqnarray}
denote the fraction of the proton's spin carried by the light quarks and antiquarks with flavor 
$q=u$, $d$, $s$. In the nonrelativistic limit, they are given by the matrix elements 
\begin{eqnarray}
\Delta q = 2\left< p \uparrow | s_z(q) + s_z(\bar{q}) | p \uparrow \right> .
\label{quarkspin}
\end{eqnarray}
The last term in Eq.~(\ref{jproton}) represents the contribution from orbital angular momentum  
\begin{eqnarray}
\Delta L = \sum_q \Delta L(q) = \sum_q \left< p \uparrow |  
l_z(q) + l_z(\bar{q})  | p \uparrow \right> ~.
\label{orbital}
\end{eqnarray}

In the present unquenched quark model, the $SU(3)$ flavor symmetry is satisfied by the valence 
quark configuration, but broken by the quark-antiquark pairs. In 
the unquenched calculation we use harmonic oscillator wave functions up to five oscillator shells 
for both the intermediate baryons and mesons. As mentioned in Section~\ref{ucqm}, all parameters 
were taken from the literature \cite{baryons,CR}. No attempt was made to optimize their values 
in order to improve the agreement with experimental data. 

Table~\ref{spinproton} shows that 
the inclusion of the quark-antiquark pairs has a dramatic effect on the spin content of the 
proton. Whereas in the CQM the proton spin is carried entirely by the (valence) quarks, 
it is shown in Table~\ref{spinproton} that in the unquenched calculation 67.6 \% is carried by 
the quark and antiquark spins and the remaining 32.4 \% by orbital angular momentum. 
The orbital angular momentum due to the relative motion of the 
baryon with respect to the meson accounts for 31.7 \% of the proton spin, whereas the orbitally 
excited baryons and mesons in the intermediate state only contribute 0.7 \%. Finally we note, 
that the orbital angular momentum arises almost entirely from the relative motion of the nucleon 
and $\Delta$ resonance with respect to the $\pi$-meson in the intermediate states. 
In the closure limit, all mesons (including the pion) have the same mass and their contributions 
to the orbital angular momentum average out and reduce to less than 1 \% of the proton spin.  

On the contrary, the contribution of the quark and antiquark spins to the proton spin is dominated 
by the intermediate vector mesons. Since in the case of the quark spins the convergence of the sum 
over intermediate states is slow, we carried out the sum over five oscillator shells for both the 
intermediate baryons and mesons. For each oscillator shell the sum is performed over a complete set 
of spin-flavor states. It is important to note that the contributions of the valence quark spins, 
the sea quark spins and the orbital angular momentum to the proton spin, 37.8 \%, 29.8 \% and 32.4 \%, 
respectively, are comparable in size. 

In the unquenched quark model there is a large contribution ($\sim$ 32 \%) of orbital angular 
momentum to the proton spin, while for the proton magnetic moment it is relatively small 
($\sim$ 5 \%). This can be understood qualitatively from the difference in relative signs between 
the quark and antiquark contributions in Eqs.~(\ref{magnetic}) for the magnetic moment and 
those in Eqs.~(\ref{quarkspin},\ref{orbital}) for the proton spin.  

The present results for the singlet axial charge $a_0=\Delta \Sigma$ are in qualitative agreement 
with the cloudy bag model and the Nambu-Jona-Lasinio model in which one finds $a_0=0.60$ \cite{CBM} 
and $0.56$ \cite{NJL}, respectively. The inclusion of kaon loops gives in both models a small 
value of the contribution of strange quarks, $\Delta s=-0.003$ and $-0.006$, respectively, 
in agreement with the unquenched calculations. 
Another effect of the quark-antiquark pairs is a reduction of the triplet and octet axial 
charges from their CQM values of $5/3$ and $1$ to $a_3=\Delta u - \Delta d=1.515$ and 
$a_8=\Delta u + \Delta d - 2\Delta s=0.681$, respectively. It is interesting to note that 
the ratio of these axial charges in the unquenched quark model is calculated to be $a_3/a_8=2.22$ 
which is very close to the value of $2.15$ determined from hyperon semileptonic decays, but very 
different from the naive CQM value $5/3$. 

Experimentally, the contributions of the quark spins $\Delta u$, $\Delta d$ and $\Delta s$ 
to the spin of the proton are obtained by combining data from hyperon $\beta$ decays and  
deep-inelastic lepton-nucleon scattering processes. First, the hyperon $\beta$ decays 
$n \rightarrow p + e^- + \bar{\nu}_e$ and $\Sigma^- \rightarrow n + e^- + \bar{\nu}_e$ are 
used in combination with the assumption of $SU(3)$ flavor symmetry to determine the couplings 
$F=(a_3+a_8)/4$ and $D=(3a_3-a_8)/4$. Next, $\Delta \Sigma$ can be extracted from deep-inelastic 
electron-proton scattering experiments. As a result, $\Delta q$ of the proton is given by 
\ba
(\Delta u)_{p} &=& \frac{1}{3} \left( \Delta \Sigma + 3F + D \right) ~, 
\nonumber\\
(\Delta d)_{p} &=& \frac{1}{3} \left( \Delta \Sigma - 2D \right) ~, 
\nonumber\\
(\Delta s)_{p} &=& \frac{1}{3} \left( \Delta \Sigma - 3F + D \right) ~. 
\label{proton}
\ea
The theoretical uncertainty in determining the values of $F$ and $D$ by assuming flavor 
symmetry were estimated to be of the order of 10-15 \% \cite{HDS,Manohar,Ratcliffe,Goeke}. 
It is important to keep in mind that, even though the effect of flavor symmetry breaking on 
the hyperon decays may not be so large, for other quantities like $\Delta \Sigma$ and $\Delta s$ 
it is much stronger \cite{HDS,Goeke,Savage}. 
The results of the HERMES analysis are presented in the column labeled DIS of Table~\ref{spinproton}.  
These values were obtained by combining the couplings $F=0.464$ and $D=0.806$ as determined from 
hyperon $\beta$ decays with $\Delta \Sigma=0.330$ as extracted from the first moment of the spin 
structure function $g_1^p$ \cite{hermes}. For the purpose of reference, we also present the values 
for the naive quark model (CQM) which correspond to $F=2/3$ and $D=\Delta \Sigma=1$ and 
for the Ellis-Jaffe-Sehgal analysis (EJS), in which it is assumed that there are no polarized strange 
quarks in the proton \cite{Ellis,jaffe}. In the latter case, the spin content is calculated by using 
$F$ and $D$ from hyperon $\beta$ decays and $\Delta \Sigma = 3F-D$. The remainder of the proton spin 
$1-3F+D$ is attributed to orbital angular momentum \cite{Sehgal}.

The importance of orbital angular momentum to the proton spin was discussed many years ago 
by Sehgal \cite{Sehgal} in the context of the quark-parton model. Table~\ref{spinproton} shows, 
that the results of the unquenched quark model are similar to those of the EJS analysis. 
More recently, Myhrer and Thomas emphasized the importance of spin and orbital angular 
momentum in the proton in the bag model \cite{thomas_spin} and discussed three effects that can 
convert quark spin into orbital angular momentum: the relativistic motion of the valence quarks, 
the one-gluon exchange corrections and the pion cloud of the nucleon. The contribution of the 
quark spins was estimated in a qualitative way to be in the range $0.35 < \Delta \Sigma < 0.40$.  

\subsection{$\Lambda$ spin}

\begin{table}[tb]
\centering
\caption[]{As Fig.~\ref{spinproton}, but for the $\Lambda$ hyperon.}
\label{spinlambda}
\begin{tabular}{ccrrrrr}
\noalign{\smallskip}
\hline
\noalign{\smallskip}
& & & & \multicolumn{3}{c}{UCQM} \\
$\Lambda$ & CQM & EJS & DIS & val & sea & total \\
\noalign{\smallskip}
\hline
\noalign{\smallskip}
$\Delta u$      & 0 & --0.073 & --0.159 & 0.000 & --0.055 & --0.055 \\
$\Delta d$      & 0 & --0.073 & --0.159 & 0.000 & --0.055 & --0.055 \\
$\Delta s$      & 1 &   0.733 &   0.647 & 0.422 &   0.539 &   0.961 \\
$\Delta \Sigma$ & 1 &   0.586 &   0.330 & 0.422 &   0.429 &   0.851 \\
$2 \Delta L$    & 0 &   0.414 &         & 0.000 &   0.149 &   0.149 \\
$2J$            & 1 &   1.000 &         & 0.422 &   0.578 &   1.000 \\
\noalign{\smallskip}
\hline
\end{tabular}
\end{table}

The recent studies of the spin structure of the proton have raised a lot of questions  
about the importance of valence and sea quarks, gluons and orbital angular momentum. 
In this respect it is interesting to investigate the spin structure of other hadrons.   
The $\Lambda$ hyperon is of special interest, since its polarization can be measured 
from the nonleptonic decay $\Lambda \rightarrow p \pi$ \cite{jaffe}. In addition, 
in the naive CQM its spin content resides entirely on the strange quark, 
$(\Delta u)_{\Lambda}=(\Delta d)_{\Lambda}=0$ and $(\Delta s)_{\Lambda}=1$, which 
makes it a clean example to study the spin structure of baryons. An investigation 
of the spin structure of the $\Lambda$ hyperon is not only interesting in its own right, 
but also may shed light on the spin crisis of the proton.  

Table~\ref{spinlambda} shows that the unquenched quark model gives rise to a negatively 
polarized sea of up and down quarks. The contribution of quark spins for the $\Lambda$ 
is found to be larger than that for the proton, $(\Delta \Sigma)_{\Lambda} > (\Delta \Sigma)_{p}$. 

It is interesting to compare the unquenched results with those of some previous analyses. 
In most other studies one had to make additional assumptions about the sea quarks in order to 
get an estimate of the spin content of the $\Lambda$ hyperon in most. Under the assumption 
of $SU(3)$ flavor symmetry, the spin content of the octet baryons can be expressed in terms 
of that of the proton as \cite{cabibbo,jaffe} 
\ba
(\Delta u)_{\Lambda} = (\Delta d)_{\Lambda} &=& 
\frac{1}{6} \left( \Delta u + 4\Delta d + \Delta s \right)_{p} 
\nonumber\\
&=& \frac{1}{3} \left( \Delta \Sigma - D \right) ~, 
\nonumber\\
(\Delta s)_{\Lambda} &=& 
\frac{1}{3} \left( 2\Delta u - \Delta d + 2\Delta s \right)_{p} 
\nonumber\\
&=& \frac{1}{3} \left( \Delta \Sigma + 2D \right) ~,
\label{lambda}
\ea
In this case, it is assumed that both the valence and sea quarks are related by $SU(3)$ 
flavor symmetry. As an example of this procedure, we present in Table~\ref{spinlambda} 
the results for the spin content of the $\Lambda$ hyperon in the Ellis-Jaffe-Sehgal analysis  
(EJS) and another one based on the DIS results for the proton (DIS). In the former, it is 
found that the up and down quarks are negatively polarized and that the total contribution 
from the quarks and antiquarks to the $\Lambda$ spin is reduced to $\Delta \Sigma = 0.586$ 
\cite{jaffe}. An analysis of the experimental DIS data for the proton \cite{hermes,compass} 
in combination with Eq.~(\ref{lambda}) shows that the strange quarks (and antiquarks) 
carry about 65 \% of the $\Lambda$ spin, while the up and down quarks (and antiquarks) 
account for a negative polarization of --32 \%. The negative polarization of the up and 
down quarks is confirmed by different theoretical studies, such as the chiral quark-soliton 
model \cite{Goeke}, lattice QCD \cite{lattice} and QCD sum rules \cite{oka}. It has been 
pointed out, that $SU(3)$ symmetry breaking effects in hyperon $\beta$ decays may reduce 
the negative polarization \cite{Goeke,yang}. 

Another assumption about the sea sometimes used in the literature is that the sea polarization 
is the same for all octet baryons, whereas the valence quarks are related by $SU(3)$ symmetry 
\cite{jaffe,lipkin}. However, experimental information on the violation of the Gottfried sum 
rule \cite{gsr} and the suppression of the polarized strange quark momentum contribution with 
respect to that of the nonstrange quarks \cite{neutrino}, shows that the sea quark distributions 
depend on the valence quark content in a nontrivial manner.

In the unquenched quark model there is no need to make additional assumptions about the nature 
of the sea. The valence quarks are related by $SU(3)$ flavor symmetry, but the flavor symmetry is 
broken by the the sea quarks (see Eq.~(\ref{baryonwf})). Therefore, the $SU(3)$ flavor symmetry 
relations in Eq.~(\ref{lambda}) do not hold in the unquenched calculations. Table~\ref{spinlambda} 
shows that, just as for EJS and DIS, the unquenched quark model gives rise to a negatively polarized 
sea of up and down quarks, but its results are a lot closer to the CQM values than those of EJS and DIS. 
The present analysis of the spin content of the proton and the $\Lambda$ hyperon shows in an 
explicit way the importance of $SU(3)$ breaking effects. 

\section{Summary and conclusions}
 
There is ample experimental evidence for the importance of sea quarks in the structure of hadrons. 
In this paper, we discussed an unquenched quark model for baryons which incorporates the effects 
of quark-antiquark pairs. The quark loops are taken into account via a $^3P_0$ pair creation model.  
The ensuing unquenched quark model is valid for any baryon (or baryon resonance), includes 
all light flavors of the pairs ($u \bar{u}$, $d \bar{d}$ and $s \bar{s}$), and can be used 
for any CQM, as long as its wave functions are expressed in a harmonic oscillator basis. 

Obviously, the unquenching of the quark model has to be done in such a way that it preserves 
the phenomenological successes of the CQM. As an example, we showed that, after renormalization 
of the quark magnetic moments, the inclusion of quark-antiquark pairs does not change the good 
CQM results for the magnetic moments of the octet baryons. In a similar way, one has studied the 
effects of hadron loops on the OZI hierarchy \cite{OZI}, self-energies \cite{Capstick,swanson} 
and hybrid mixing \cite{close}. 

In an application of the unquenched quark model to the spin of the proton and the $\Lambda$ 
hyperon, it was found that the inclusion of $q \bar{q}$ pairs leads to a relatively large 
contribution of orbital angular momentum to the spin of the proton ($ \sim 32$ \%) and a 
somewhat smaller amount for $\Lambda$ ($\sim 15$ \%). The difference between these numbers 
is an indication for the breaking of $SU(3)$ flavor symmetry in the unquenched quark model. 
The valence quarks are related by flavor symmetry, but the contribution of the sea quarks 
is determined by the $^3P_0$ coupling between the valence quarks and the higher Fock states 
without any additional assumption. The contribution of strange quarks to the proton spin is 
found to be very small, in agreement with results in the cloudy bag model and the NJL model. 
The relative contribution of up and down quarks $\Delta u/\Delta d$ is reduced from $-4$ in 
the CQM to $-2.6$. For the $\Lambda$ hyperon we found a small contribution of a negatively 
polarized sea of up and down quarks, in qualitative agreement with other studies. The spin 
content of $\Lambda$ is dominated by the strange quark spins. The results of the unquenched 
quark model for the spin content of $\Lambda$ are much closer to the CQM values than that of 
the proton. In order to be able to make a more detailed comparison with experimental data, 
one has to include the effects of relativity and evolve the scale dependent quantities to 
the experimental scale. The present results represent a first step. 
Relativistic calculations are underway in front form and point form dynamics \cite{polyzou}. 
 
The sum over intermediate baryon-meson states is carried out explicitly and includes all 
possible intermediate states: singlet, octet and decuplet baryons and pseudoscalar and vector 
mesons as well as their orbital excitations up to any oscillator shell. 
The convergence of the sum depends on the quantity 
one is interested in. For the orbital angular momentum, the convergence is very rapid, since 
the sum is dominated by the contribution of the pions. On the other hand, for the quark spins 
the sum over intermediate states is dominated by the contribution of the vector mesons and 
many oscillator shells have to be included before reaching convergence. 

The main idea of this paper was to present an unquenched quark model in which the effects of 
quark-antiquark pairs are introduced explicitly, and which offers the possibility to study 
the importance of $q \bar{q}$ pairs in hadrons in a systematic and unified way. 
To the best of our knowledge, these are the first explicit calculations of the sea 
contributions in the quark model. The present results for the magnetic moments and the spin 
content of octet baryons in combination with preliminary results for the flavor asymmetry of 
the nucleon \cite{ucqm2} are very promising and encouraging. We believe that the inclusion of 
the effects of quark-antiquark pairs in a general and consistent way, as suggested here, may 
provide a major improvement to the constituent quark model which increases considerably its 
range of applicability. 

\section*{Acknowledgments}

We thank the late Nathan Isgur for interesting discussions and encouragement in the early stages of 
this work, and Mauro Giannini for stimulating discussions and his continuous interest. 
This work was supported in part by a grant from INFN, Italy  
and in part by grant no. 78833 from CONACYT, Mexico.

\end{document}